\documentclass[sigconf]{acmart}
\usepackage{enumitem}
\usepackage{graphicx}
\usepackage{subfigure}
\AtBeginDocument{%
  }

\copyrightyear{2025}
\acmYear{2025}
\setcopyright{rightsretained}
\acmConference[RecSys '25]{Proceedings of the Nineteenth ACM Conference on
Recommender Systems}{September 22--26, 2025}{Prague, Czech Republic}
\acmBooktitle{Proceedings of the Nineteenth ACM Conference on Recommender
Systems (RecSys '25), September 22--26, 2025, Prague, Czech
Republic}\acmDOI{10.1145/3705328.3748107}
\acmISBN{979-8-4007-1364-4/2025/09}

\begin{document}

\title{User Long-Term Multi-Interest Retrieval Model for Recommendation}

\author{Yue Meng}
\email{mengyue.meng@taobao.com}
\author{Cheng Guo}
\email{mike.gc@taobao.com}
\author{Xiaohui Hu}
\email{yixiao.hxh@taobao.com}
\affiliation{%
  \institution{Taobao \& Tmall Group of Alibaba}
  \city{Beijing}
  \country{China}
}

\author{Honghu Deng}
\email{dhh23@mails.tsinghua.edu.cn}
\affiliation{%
  \institution{Tsinghua University}
  \city{Beijing}
  \country{China}
}

\author{Yi Cao}
\authornote{Corresponding author}
\email{dylan.cy@taobao.com}
\author{Tong Liu}
\email{yingmu@taobao.com}
\author{Bo Zheng}
\email{bozheng@alibaba-inc.com}
\affiliation{%
  \institution{Taobao \& Tmall Group of Alibaba}
  \city{Hangzhou}
  \country{China}
}

\renewcommand{\shortauthors}{Meng et al.}

% 是否要给出具体的数字

\begin{abstract}
    User behavior sequence modeling, which captures user interest from rich historical interactions, is pivotal for industrial recommendation systems. Despite breakthroughs in ranking-stage models capable of leveraging ultra-long behavior sequences with length scaling up to thousands, existing retrieval models remain constrained to sequences of hundreds of behaviors due to two main challenges. One is strict latency budget imposed by real-time service over large-scale candidate pool. The other is the absence of target-aware mechanisms and cross-interaction architectures, which prevent utilizing ranking-like techniques to simplify long sequence modeling. To address these limitations, we propose a new framework named User Long-term Multi-Interest Retrieval Model(ULIM), which enables thousand-scale behavior modeling in retrieval stages. ULIM includes two novel components: 1)Category-Aware Hierarchical Dual-Interest Learning partitions long behavior sequences into multiple category-aware subsequences representing multi-interest and jointly optimizes long-term and short-term interests within specific interest cluster. 2)Pointer-Enhanced Cascaded Category-to-Item Retrieval introduces Pointer-Generator Interest Network(PGIN) for next-category prediction, followed by next-item retrieval upon the top-K predicted categories. Comprehensive experiments on Taobao dataset show that ULIM achieves substantial improvement over state-of-the-art methods, and brings 5.54\% clicks, 11.01\% orders and 4.03\% GMV lift for Taobaomiaosha, a notable mini-app of Taobao.

\end{abstract}





\maketitle

\section{Introduction}

Powered by the exponential growth of historical user behavior data, modeling user interests has become fundamental to industrial recommendation systems\cite{DBLP:conf/recsys/CovingtonAS16, DBLP:conf/kdd/PiBZZG19}.
While sequential recommendation—such as DIN\cite{DBLP:conf/kdd/ZhouZSFZMYJLG18}, SIM\cite{DBLP:conf/cikm/PiZZWRFZG20}, ETA\cite{DBLP:journals/corr/abs-2108-04468}—scale interest representations to thousands of actions, they have largely focused on the ranking\cite{DBLP:conf/wsdm/ChenXZT0QZ18, DBLP:conf/cikm/LvJYSLYN19}. 
In contrast, retrieval models typically only consider tens of behaviors, limiting consistency with downstream rankings.

The scarcity of research on long-sequence modeling\cite{DBLP:conf/kdd/LiLJLYZWM21, DBLP:conf/sigir/ZhangWZTJXYY20} stems from two key challenges. 
First, unlike the few thousand shortlist of ranking, retrieval must scan tens of millions of candidates, making the compute and latency requirements prohibitively high. Second, conventional retrieval architectures, framed as next-item prediction, lack target-aware interactions\cite{DBLP:conf/icml/ZhaiLLWLCGGGHLS24} and cannot exploit hierarchical simplifications such as category-based searching\cite{DBLP:conf/cikm/PiZZWRFZG20}.  

In addition, multi-interest modeling should be considered in retrieval stages\cite{DBLP:conf/kdd/CenZZZYT20, DBLP:conf/wsdm/TanZYLZYH21, DBLP:conf/sigir/LiQPQDW19, DBLP:conf/recsys/XieGZYHKWK23}. 
MIND\cite{DBLP:conf/cikm/LiLWXZHKCLL19} employs a capsule network\cite{DBLP:conf/icann/HintonKW11} with dynamic routing\cite{DBLP:conf/nips/SabourFH17} to cluster user behaviors into distinct interest representations. 
% Single-interest modeling in inner-product-based retrieval would cause gradient conflict that user embeddings are pulled by diverse item embeddings, ultimately degrading the embedding quality.
Single-interest, inner-product retrieval induces gradient conflicts: embedding updates drawn toward diverse items degrade overall representation quality.

To address these issues, we propose the \textbf{User Long-term Multi-Interest Retrieval Model (ULIM)}, which integrates thousands of long behavior sequences into the retrieval stage via:
\begin{enumerate}[label=\roman*)]
    \item \emph{Category-Aware Hierarchical Dual-Interest Learning} (training): partitioning long behavior sequences into category-aware subsequences and jointly optimizing long‑ and short‑term multi‑interest representations.
    \item \emph{Pointer‑Enhanced Cascaded Category‑to‑Item Retrieval} (serving): a pointer‑generator network predicts the top‑$K$ interest categories, enabling parallel and category‑specific item retrieval that drastically reduces online computation.
\end{enumerate}

In summary, the main contributions of this paper are as follows.
\begin{itemize}
    \item We propose ULIM, a novel model that integrates long behavior sequences into retrieval-stage sequential modeling.
    \item We propose Category‑Aware Hierarchical Dual‑Interest Learning, which leverages category clustering and a revised training objective to optimize long‑ and short‑term interests.
    \item We design a Pointer-Enhanced Cascaded Category-to-Item Retrieval mechanism that predicts interest categories with PGIN and retrieves items within those categories in parallel.
    \item Comprehensive experiments on large-scale Taobao datasets show that ULIM achieves state-of-the-art performance for Taobaomiaosha, a notable mini-app of Taobao.
\end{itemize}

%  篇幅有限，先删
% \section{PROBLEM FORMULA}

% A typical recommendation system usually consists of three main stages, i.e., matching, ranking and re-ranking\cite{DBLP:conf/cikm/WilhelmRBJCG18}. Denote the result from ranking stage as $I = \left\{i_1, i_2, ..., i_{l_s}\right\}$ with length $l_s$. Re-ranking aims to reordering $I$ so as to return an optimal ordered list $R^* = \left\{i^*_1, i^*_2, ..., i^*_{l_o}\right\}$, where $l_o$ refers to actual length of recommendation results. In practice, $l_s$ is always larger than $l_o$.

% Introducing the context $C$, we can formally express re-ranking that integrates multi-objective optimization as follows: 
% \begin{equation}
% R^* = argmax_{R'} [ \alpha * V_{click}(R', C) + \beta * V_{conversion}(R', C) + \gamma * V_{GMV}(R', C)] 
% \end{equation}
% where $R'$ is a potential reordering result. $V_{click}$, $V_{conversion}$ and $V_{GMV}$ represent the measurement of click value, conversion value and GMV value at list-level respectively. $\alpha$ and $\beta$ represent the trade-offs at the Pareto boundaries for commercial purposes. List value estimation with context is structured as a sequential incremental value accumulation approach. For example, $V_{click}$ is expressed as:
% \begin{equation}
% V_{click}(R', C) = v_{click}(i'_1) + v_{click}(i'_2 | [i'_1]) + ... + v_{click}(i'_{l_o} | [i'_1, ..., i'_{l_o-1}])
% \end{equation}
% where $v_{click}$ refers to click value at item-level.

\section{The Proposed Model}
% Sec.3.4 outlines how we deployed SORT-Gen at Taobao.
% \begin{figure*}[htbp]
% \centering
% \includegraphics[width=\linewidth]{MMR.pdf}
% \caption{An overview of the SORT-Gen framework.}
% \label{fig1}
% \end{figure*}

\subsection{Category-Aware Hierarchical Dual-Interest Learning}
\begin{figure}[t]
  \centering
  \includegraphics[width=\linewidth]{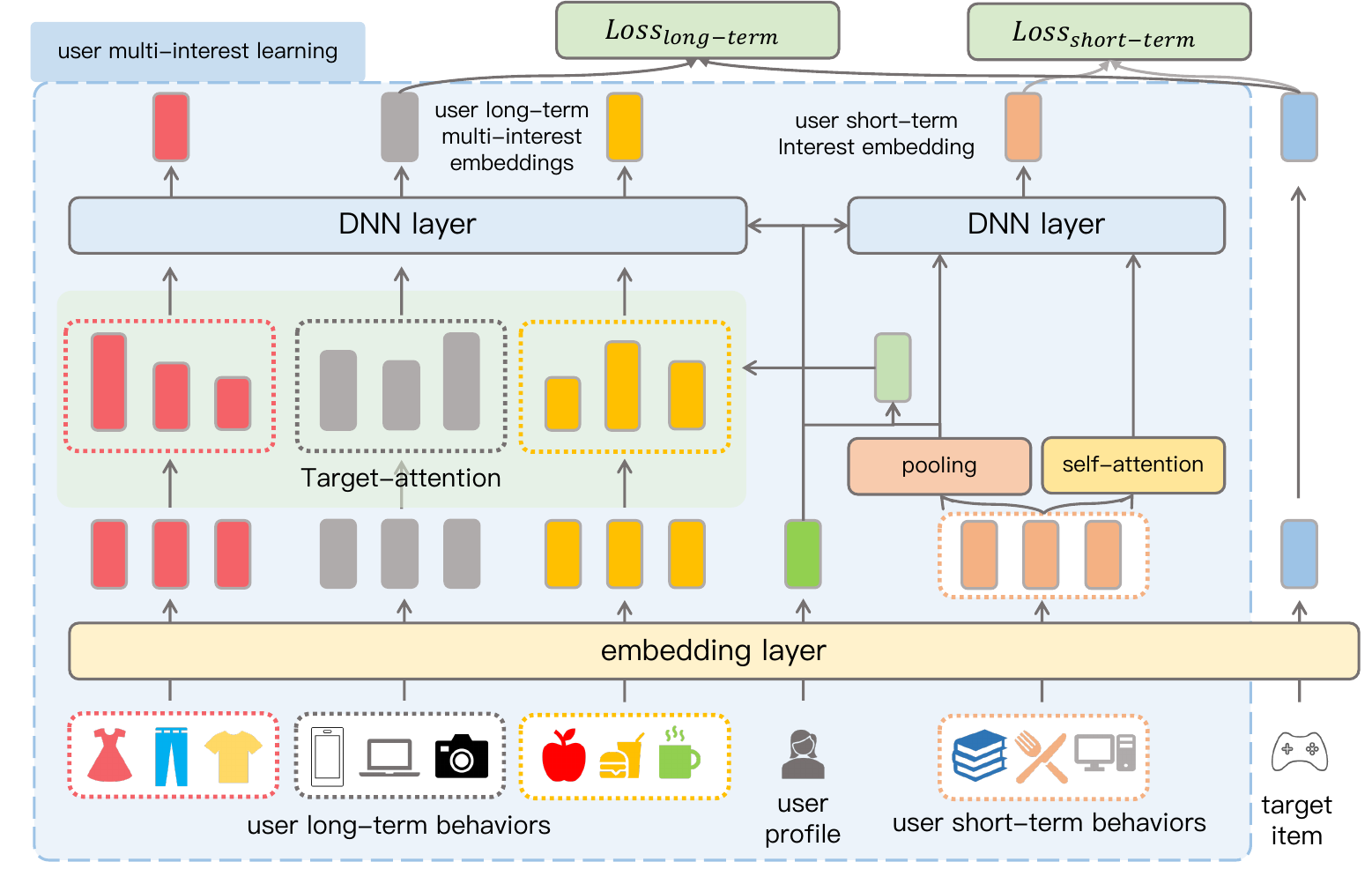}
  \vspace{-0.70 cm}
  \caption{The Framework of Category-Aware Hierarchical Dual-Interest Learning}
  \label{fig:ulim-1}
  \vspace{-0.5 cm}
\end{figure}

\subsubsection{Granularity-Aligned Category Clustering}
% \subsubsection{Pre-Grouping and Training Objective Redefinition}
We partition the raw sequence into category‐aware subsequences, producing interest clusters that reduce complexity from \(O(L)\) to \(O(L/N)\) (with \(N\) categories) and support multi‐interest modeling. 
Importantly, we align the category granularity with that of the ranking stage to maintain a consistent feature space across retrieval and ranking.

\subsubsection{Training Objective Redefinition}
We redefine the training objective of retrieval models. ULIM predicts the click probabilities within each interest cluster, rather than over the entire candidate pool.
For any training sample, the positive sample strictly matches the category of the corresponding subsequence, and all negative samples are drawn from the same category subspace. 
By pre-grouping data into category-homogeneous batches and sharing negatives within each batch, we block category leakage and prevent trivial classification, training collapse, and inflated offline metrics.

\subsubsection{Model Structure}
% Fig. \ref{fig:ulim-1} illustrates the framework of Category-Aware Hierarchical Dual-Interest Learning, which employs two parallel sequence encoders with cross-sequence interaction to capture both short- and long-term user interests and an item tower for generating item embeddings.

% Short-Term Interest Encoder processes a user’s recent behavior sequence (up to 100 actions) using Multi-head Self-Attention\cite{DBLP:conf/nips/VaswaniSPUJGKP17} followed by average pooling. 
% This encoder can directly reuse existing architectures (e.g., YouTube DNN\cite{DBLP:conf/recsys/CovingtonAS16} or MIND\cite{DBLP:conf/cikm/LiLWXZHKCLL19}).

% Long-Term Interest Encoder obtains specific category-aware subsequences through Granularity-Aligned Category Clustering, and then uses the Target-Attention mechanism to generate the user's long-term interest embedding. 
% The Target-Attention mechanism realizes the interaction between long-term and short-term behaviors. 
% In order to ensure the parallelization of the two sequence encoders, the Target-Attention mechanism only uses the pooling results of short-term user behaviors.

% During training, each positive sample activates exactly one category-aware subsequence. At serving time, each category-aware subsequence can be processed in parallel using Long-Term Interest Encoder to obtain $K$ user long-term multi-interest embeddings.
%%%%%%%%%%%%%%%%%%%%%%
The Category-Aware Hierarchical Dual-Interest Learning framework (Fig. \ref{fig:ulim-1}) incorporates two sequence encoders in parallel to capture long- and short-term user interests, along with an item tower for item embedding generation.

The Short-Term Interest Encoder processes the user's recent behavior sequence (up to 100 actions) using Multi-head Self-Attention (MHSA)\cite{DBLP:conf/nips/VaswaniSPUJGKP17} followed by average pooling.

The Long-Term Interest Encoder extracts category-aware subsequences via Granularity-Aligned Category Clustering, then performs Target-Attention using short-term pooling results as the query to generate the user's long-term interest embedding.During training, each positive sample activates a single category-aware subsequence. During serving, multiple subsequences can be processed in parallel to obtain $K$ long-term multi-interest embeddings.

\begin{figure}[t]
  \centering
  \includegraphics[width=\linewidth]{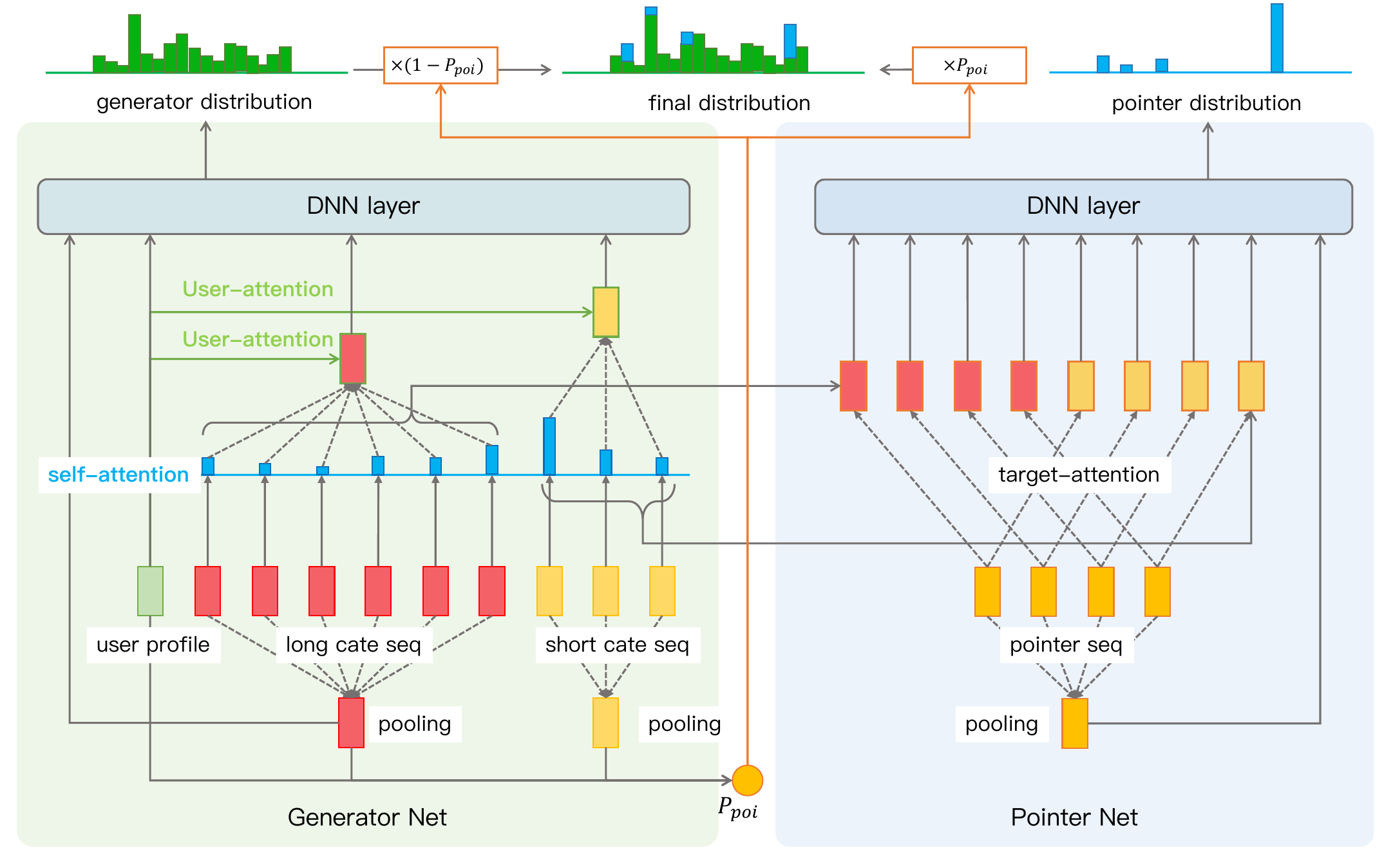}
  \vspace{-0.70 cm}
  \caption{Pointer-Generator Interest Network}
  \label{fig:module2}
  \vspace{-0.5 cm}
\end{figure}

\subsubsection{Loss Function}
To ensure discriminative and distinct long- and short-term interest embeddings, we optimize the following composite loss:
\begin{equation}
\mathcal{L} = \alpha \mathcal{L}_l + \beta \mathcal{L}_s,
\end{equation}
where \( \mathcal{L}_l \) and \( \mathcal{L}_s \) are sampled softmax objectives for long- and short-term embeddings, specifically:
\begin{equation}
\begin{aligned}
\mathcal{L}_l &= -\sum_{(u, i) \in \mathcal{D}} \log \left[ \frac{\exp \left( \mathbf{v}_u^{\text{(l)}} \top \mathbf{\vec{e}}_i \right)}{\sum_{j \in \mathcal{I}_{\text{neg}}} \exp \left( \mathbf{v}_u^{\text{(l)}} \top \mathbf{\vec{e}}_j \right)} \right], \\
\mathcal{L}_s &= -\sum_{(u, i) \in \mathcal{D}} \log \left[ \frac{\exp \left( \mathbf{v}_u^{\text{(s)}} \top \mathbf{\vec{e}}_i \right)}{\sum_{j \in \mathcal{I}_{\text{neg}}} \exp \left( \mathbf{v}_u^{\text{(s)}} \top \mathbf{\vec{e}}_j \right)} \right], \\
% \mathcal{L}_{\text{gap}} &= \frac{1}{m(m-1)} \sum_{i \neq j} v_i \cdot v_j,
\end{aligned}
\end{equation}
where \( \mathcal{D} \) is the training dataset, \( i \) is an item clicked by the user, \( \mathbf{\vec{e}}_i \) is the item embedding, \( \mathbf{v}_u^{\text{(l)}} \) and \( \mathbf{v}_u^{\text{(s)}} \) are the user’s long- and short-term interest embeddings.

\subsection{Pointer-Enhanced Cascaded Category-to-Item Retrieval}

To align offline training with online serving, we propose Pointer-Enhanced Cascaded Category-to-Item Retrieval, which is a two-stage hierarchical online retrieval framework. 
First, the Pointer-Generator Interest Network (PGIN) predicts a distribution over item categories. 
Second, we perform ANN\cite{DBLP:journals/pami/JegouDS11, DBLP:journals/pami/MalkovY20} searches in parallel using all long- and short-term user interest embeddings.

% To align offline training with online serving, we introduce Pointer‑Enhanced Cascaded Category‑to‑Item Retrieval, a two‑stage framework:
% first, the Pointer‑Generator Interest Network (PGIN) predicts a distribution over item categories;
% second, we perform ANN\cite{DBLP:journals/pami/JegouDS11, DBLP:journals/pami/MalkovY20} searches in parallel using both long- and short-term user interest embeddings.

\subsubsection{Pointer-Generator Interest Network}
We propose PGIN, an enhanced Pointer‑Generator architecture for user interest prediction (Fig. \ref{fig:module2}). 
In PGIN, category prediction is cast as multiclass classification via Pointer-Net\cite{DBLP:conf/nips/VinyalsFJ15} and Generator-Net.

The Pointer-Net ingests a deduplicated, time‑ordered sequence of merged long‑ and short‑term category histories and applies Target-Attention to extract multiscale category features. 
Then a backward projection maps the outputs to the full category space. 

The Generator‑Net encodes raw long‑ and short‑term behavior sequences via MHSA and pooling, fuses these with user features through Target‑Attention, and outputs category probabilities via a stacked MLP. 
A learned gating network blends these outputs:  
\[
\widehat{y} \;=\; P_{\text{poi}}\,y_{\text{poi}} \;+\; (1 - P_{\text{poi}})\,y_{\text{gen}},
\]  
trained with cross-entropy over the true category labels.

\subsubsection{Category-Constrained Retrieval}
% Building upon PGIN, we obtain user's estimated click probability over all categories and select top-K categories. 
% We develop a multi-vector parallel retrieval engine that maintains category-specific item indices and executes Approximate Nearest Neighbor(ANN)\cite{DBLP:journals/pami/JegouDS11, DBLP:journals/pami/MalkovY20} searches for all K+1 user interest embeddings simultaneously. During online retrieval:
% \begin{enumerate}[label=\roman*)]
% \item For each predicted category, we extract the category-aware subsequence from user long behavior sequence, ultimately generating K user long-term multi-interest embeddings. 
% \item We derive the user short-term interest embedding from user short behavior sequence.
% \end{enumerate}
% All these K+1 vectors engage in parallel retrieval. 
% To maintain offline-online consistency, user long-term multi-interest embeddings can only retrieve items exclusively from its corresponding category’s candidates, whereas user short-term interest embedding searches the entire candidate pool. 
% Then result fusion that merging retrieved items from all embeddings follows. 
% This constraint aligns with our redefined training objective, effectively preventing distribution shift between offline training and online serving.
%%%%%%%%%%%
% Using the top-K categories predicted by PGIN, we perform parallel ANN searches on \(K+1\) user embeddings: one short-term interest embedding over the full candidate pool and \(K\) long-term interest embeddings restricted to their respective category indices. At serving time:
Using the top-K categories predicted by PGIN, we perform parallel ANN searches on \(K+1\) interest embeddings. At serving time:
\begin{enumerate}[label=\roman*)]
    \item For each predicted category, we extract the category-aware subsequence from the long-term behavior sequence, ultimately generating $K$ long-term interest embeddings. 
    \item Compute the embedding of short-term interest from the short-term behavior sequence.
\end{enumerate}
To maintain offline-online consistency, user long-term multi-interest embeddings can only retrieve items exclusively from its corresponding category’s candidates, whereas user short-term interest embedding searches the entire candidate pool.
This constraint aligns with the redefined training objective and avoids distribution shift between offline training and online serving.

\section{Experiments}

% 2.1 对比其他召回模型的离线hitrate比较
% 2.2 线上inference time，和mind、双塔对比；相同检索数量等配置下
% 2.3 线上实际效果
% # 2.4 在线消融实验：场域&全域，类目选择策略* png和规则类对比，类目选择粒度
% 2.5 离线消融实验：模型结构离线hirtate比较，各个模块消融
% 2.6 超参调优：兴趣数量，K的设计和讨论，超参数敏感性
% 2.7 在线pvr可以不写

% To evaluate the effectiveness of our proposed SORT-Gen, we conduct extensive experiments on online deployment. There are three research questions investigated in these experiments:

% \begin{itemize}
% \item \textbf{Q1}:How does SORT-Gen perform in real-life recommendation systems?
% \item \textbf{Q2}:How much impact do different modules of SORT-Gen have on the final results?
% \item \textbf{Q3}:How does SORT-Gen specifically work in real-life recommendation systems to achieve improvement?
% \end{itemize}

\subsection{Offline Experiments}
\subsubsection{Experiment Setup}

\begin{table}
  \caption{Offline experiments performance on Taobao dataset.}
  \label{tab:freq}
  \begin{tabular}{p{3cm}ccc}
    \toprule
    Method & HR@500 & HR@1000 & HR@2000\\
    \midrule
    Youtube DNN Variant & 4.95\% & 9.53\% & 14.93\%\\
    MIND Variant & 5.03\% & 9.66\% & 15.15\%\\
    % Optimized MIND & 13.09 & 18.93 & 24.25\\
    \textbf{ULIM} & \textbf{6.02\%} & \textbf{10.76\%} & \textbf{16.55\%}\\
    % $\pi$ & 1 in 5& Common in math\\
    % \$ & 4 in 5 & Used in business\\
    % $\Psi^2_1$ & 1 in 40,000& Unexplained usage\\
  \bottomrule
\end{tabular}
\end{table}

We conduct experiments on Taobao dataset including over tens of millions of positive samples daily. We compare our proposed model ULIM with state-of-the-art retrieval methods adapted for our scenarios including Youtube DNN\cite{DBLP:conf/recsys/CovingtonAS16} and MIND\cite{DBLP:conf/cikm/LiLWXZHKCLL19}. These baselines were optimized through data migration and multi-objective optimization. Notably, the enhanced Mind variant has already been deployed as one of the online retrieval channels.
%and optimized MIND, which utilizes the entire space samples by transfer learning. Optimized MIND is the best model before and has been deployed as one of online retrieval channels in various scenarios of Taobao App. % In the offline experiments, we set the predicted categories number K as 10. 
In our experiments, ULIM utilizes two years of user historical behavior. To ensure fairness, other parameters such as embedding dimension, the number of retrieved items and optimizer are all kept consistent. We use Hit Rate(HR), which has been widely used in previous work\cite{DBLP:conf/wsdm/ChenXZT0QZ18}\cite{DBLP:conf/cikm/Karypis01}, to evaluate the offline performance.

\subsubsection{Offline Experiment Results}
Table 1 summarizes the performance of ULIM as well as baselines on Taobao's large-scale industrial dataset. Clearly, ULIM outperforms others by a wide margin on the evaluation criteria. Results show that the method utilizing user long behavior sequences has proved to be more effective.

% In addition to the efficiency metrics mentioned above, we also need to focus on user experience-related metrics, which are presented in Table 2.
\subsubsection{Parameter Sensitivity}

% 补充个图
We investigate the sensitivity of the number of selected categories $K$, which also represents the number of interest clusters. Fig. 3 reports the results in terms of HR. It demonstrates that greater $K$ achieves better offline HR@2000 while exhibiting marginal effect. Besides, greater $K$ would introduce more online Response Time(RT). Therefore, the optimal selection of K is heavily influenced by the practical characteristic of different scenarios and should be determined experimentally.

\begin{figure}[h]
  \centering
  \includegraphics[width=0.6\linewidth]{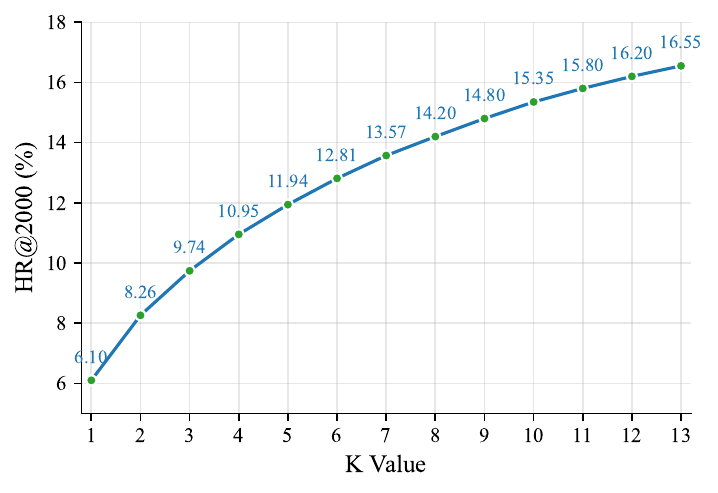}
  \caption{Sensitivity of ULIM towards K on Taobao}
\end{figure}

% 考虑补充负采样的敏感度

\subsection{Online Experiments}

Online A/B experiments have been launched on Taobaomiaosha, a notable mini-app of Taobao, over three weeks. Results show that ULIM brings 5.54\% clicks, 11.01\% orders, 4.03\% GMV lift. We integrated ULIM as in independent retrieval channel into our current multi-channel retrieval, contributing incremental items for the ranking stage. The overall recommendation system RT increased by approximately 15ms, while achieving significant improvements across nearly all metrics, delivering substantial industrial value and commercial returns.

% \begin{table}
%   \caption{Online experiments results on Taobaomiaosha.}
%   \label{tab:freq}
%   \begin{tabular}{cccc}
%     \toprule
%     CLICK & ORDER & GMV & STAY\_TIME\\
%     \midrule
%     +5.54\% & +11.01\% & +4.03\% & +5.40\%\\
%     % $\pi$ & 1 in 5& Common in math\\
%     % \$ & 4 in 5 & Used in business\\
%     % $\Psi^2_1$ & 1 in 40,000& Unexplained usage\\
%   \bottomrule
% \end{tabular}
% \end{table}

\subsection{Ablation Study}
We validate the effectiveness of different modules in ULIM through ablation studies. Three variants are proposed from different prospective. Table 2 quantifies the results. ULIM-half-sequence tests the effect of halving the length of user long behavior sequences in the feature. 
%ULIM-add-loss is used to verify the effectiveness of dual interest loss. 
ULIM-self-attention replaces Target-Attention with Self-Attention in long sequence modeling, which prevents cross-sequence interaction. Obviously, ULIM significantly outperforms the other three variants on Taobao dataset.
\begin{table}
  \caption{Results of ablation study.}
  \label{tab:freq}
  \begin{tabular}{p{3cm}ccc}
    \toprule
    Method & HR@500 & HR@1000 & HR@2000\\
    \midrule
    ULIM-half-sequence & 4.75\% & 8.03\% & 13.36\%\\
    % ULIM-add-loss & 5.88\% & 9.86\% & 16.01\%\\
    ULIM-self-attention & 5.39\% & 9.70\% & 15.22\%\\
    % Optimized MIND & 13.09 & 18.93 & 24.25\\
    \textbf{ULIM} & \textbf{6.02\%} & \textbf{10.76\%} & \textbf{16.55\%}\\
    % $\pi$ & 1 in 5& Common in math\\
    % \$ & 4 in 5 & Used in business\\
    % $\Psi^2_1$ & 1 in 40,000& Unexplained usage\\
  \bottomrule
\end{tabular}
\end{table}

\section{Conclusion}

We present ULIM, a novel framework enabling user long behavior modeling in retrieval stages. By introducing Category-Aware Hierarchical Dual-Interest Learning and Pointer-Enhanced Cascaded Retrieval, ULIM effectively addresses latency constraints while capturing user long-term and short-term multiple interests, which brides the gap between retrieval and ranking stags and provides new insights for industrial recommendation systems. Empirical experiments on Taobao demonstrate its superiority.
   
\bibliographystyle{ACM-Reference-Format}
\bibliography{sample-base}

\section*{Main Author Bio}
\textbf{Yue Meng} is a researcher in the Department of Search and
Recommendation at Taobao \& Tmall Group of Alibaba. He received his master degree from Peking University. His research focuses on recommendation system.

\noindent \textbf{Cheng Guo} is a researcher in the Department of Search and
Recommendation at Taobao \& Tmall Group. He received his master degree from Tsinghua University, specializing in information retrieval at THUIR (Information Retrieval Lab at Tsinghua University).

\noindent \textbf{Yi Cao} is the leader of Marketing Algorithm in the Department of Search and Recommendation at Taobao \& Tmall Group of Alibaba. He received his master degree from Zhejiang University. His research focuses on recommendation system.

\end{document}